\documentclass[12pt]{article}
\usepackage{graphicx}
\newlength{\M}\M .9\textwidth
\newcommand{\Vol} [1]{{\bf #1}}
\newcommand{\Mprb} [1]{Phys. Rev. B \Vol{#1}}
\newcommand{\Mprl} [1]{Phys. Rev. Lett. \Vol{#1}}
\newcommand{\ird} [1]{IBM J. Res. Dev. \Vol{#1}}
\newcommand{\phl} [1]{Phys. Lett. \Vol{#1}}
\newcommand{\jmmm}[1]{J. Mag. Mag. Mat. \Vol{#1}}
\newcommand{\mrss}[1]{Mat. Res. Soc. Proc. \Vol{#1}}
\begin{document}
%
%
\title{Ab initio description of tunnel junctions}
\author{Peter Zahn and Ingrid Mertig\\
Technische Universit\"{a}t Dresden, Institut f\"{u}r Theoretische Physik,\\
D-01062 Dresden, Germany
}
\date{\today}
\maketitle
\begin{abstract}
%
Based on spin-density functional theory we calculate
the electronic structure of a tunnel junction consisting of two
magnetic Fe layers separated by an insulating vacuum barrier
selfconsistently.
For the conductance the Landauer formula is evaluated in the ballistic
limit as function of the magnetic configuration. Based on these
conductances the tunnel magnetoresistance (TMR) ratio is obtained.
We investigate the relation between TMR
ratio and spin polarization of 
the electronic structure at the metal/insulator interface.
\end{abstract}
\newpage
\section{Introduction}
The discovery of Giant MagnetoResistance \cite{baibich88,binasch89} in
1988 initiated a renewed interest in the phenomenon of spin dependent
tunneling. The first time Julliere in 1975 \cite{julliere75} succeded
to measure
the spin polarized current in a tunnel junction composed of two
ferromagnetic layers separated by an insulating barrier. Using improved
experimental techniques it is now possible to produce high quality
tunnel junctions showing a remarkable tunnel
magnetoresistance \cite{moodera95,miyazaki95}.
A widely used model to interpret the results in terms of the
polarization of the ferromagnetic leads was proposed by Julliere
\cite{julliere75}. It was based on earlier studies of
Meservey and Tedrow on tunneling between
superconductors and ferromagnets \cite{tedrow71,tedrow73,meservey94}.
The model assumed that  the tunnel current is proportional to the
product of the effective tunneling density of states of the considered
spin direction for both ferromagnetic leads. In this paper we want to
focus on the ab initio description of TMR of a Fe/Vacuum/Fe junction
using a supercell approach.
We will discuss the properties of the local and total density of states
and the influence of the finite thickness of the electrode on
current polarization and TMR ratio.

\section{Transport}
To evaluate the tunnel current in an Fe/Vacuum/Fe system we use a supercell
description. We consider unit cells with a fixed number of empty sites 
to simulate the vacuum barrier with a thickness of 0.55 nm (4 ML) and 
a variing number of Fe sites describing the electrodes.
Based on density functional theory the electronic structure of the system
was calculated selfconsistently using a Screened Korringa-Kohn-Rostoker (KKR)
method \cite{szunyogh94,zahn97}. This method is especially advantageous for
prolonged supercells and we treated supercells with up to 108 atoms for this
investigation.

We consider ballistic transport perpendicular to the layers in the
limit of zero bias. Starting from
the Landauer formula the conductance through a barrier between two
reservoirs can be expressed by the following Fermi surface integral
\cite{landauer57,buttiker88,schep95}
\begin{equation}
G({\bf n})=e^{2}\frac{A}{2}\sum
_{{\sigma},k}\;\delta(E_{k}^{\sigma}-E_{F})\;|{\bf n} \cdot {\bf v}_{k}^{\sigma}|,
\end{equation}
if we assume ideal transmission for all states at the Fermi energy
$E_F$.
$A$ denotes the cross section of the junction.
${\bf n} \cdot {\bf v}_{k}^{\sigma}$ is the component of the
Fermi velocity in current direction. The sum in Eq.(1) goes over all
states $k$ and over all spin directions $\sigma$, that means, we assume
that the electron spin is conserved during the transmission process.
The finger-print of magnetoelectronics is that the conductance changes
as a function of the magnetic configuration from antiparrallel (AP)
into parallel (P) under the influence of an external magnetic field.
For this reason, we consider the electronic structure and the conductance
of the superlattices in both magnetic configurations (see Fig. 1).
\begin{figure}
\begin{center}
 \includegraphics[clip=true,width=.6\M]{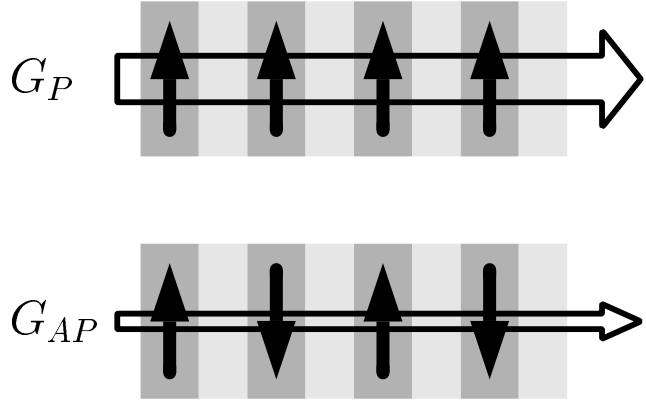}
\end{center}
 \caption{
Schematic picture of the considered structures, the directions of
the Fe magnetizations for P and AP configuration, the
direction and the usual strength of the currents
}
\end{figure}
The TMR ratio is finally defined by
\begin{equation}
TMR= 1 - \frac{G^{AP}}{G^P}\;.
\end{equation}

\section{Density of states and polarization}
Julliere \cite{julliere75} assumed that the TMR ratio is proportional
to the polarization of an effective density of states of the
ferromagnetic leads.
The question is which density is appropriate to satisfy this model.
The polarization of any density of states is defined to be
\begin{equation}
P_n=\frac{n^{\uparrow}-n^{\downarrow}}{n^{\uparrow}+n^{\downarrow}},
\end{equation}
with $n^{\sigma}$ being a spin-projected local density of states
depending on the position in the unit cell or the total density of states
for the whole system.
Results for the polarization of the local densities of states
of an Fe/Vacuum/Fe system with a vacuum barrier of 1.1 nm (8 ML)
are shown in Fig. 2.
\begin{figure}
\begin{center}
 \includegraphics[clip=true,width=\M]{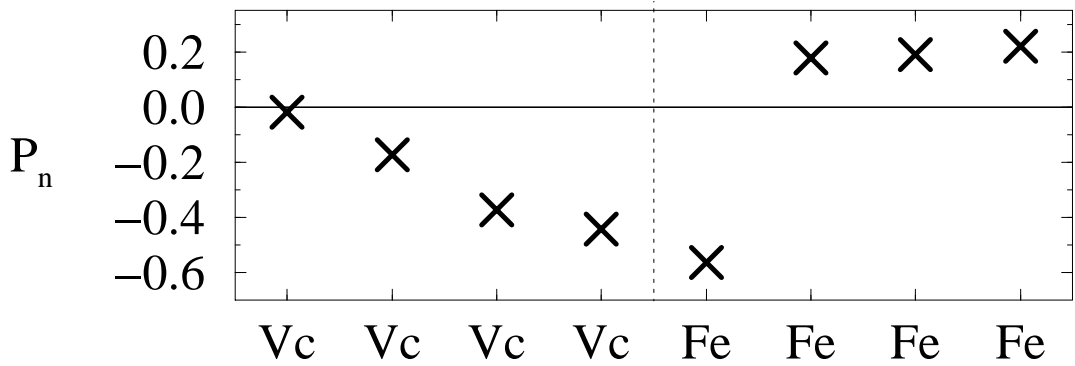}
\end{center}
 \caption{
Polarization of the local density of states at the Fermi level
for a Fe(001) surface, Vc denotes the atomic sites above the surface
}
\end{figure}

In Fig. 2 the local density of states at the Fermi level at the different sites
in the supercell is shown. The Fe polarization is positive and
bulk-like inside the Fe layers and changes sign at the Fe interface
due to the existence of surface resonances. Inside the vacuum barrier the
polarization remains negative and tends to zero for larger distances from
the surface. 

\section{TMR}
\begin{figure}
\begin{center}
 \includegraphics[clip=true,width=\M]{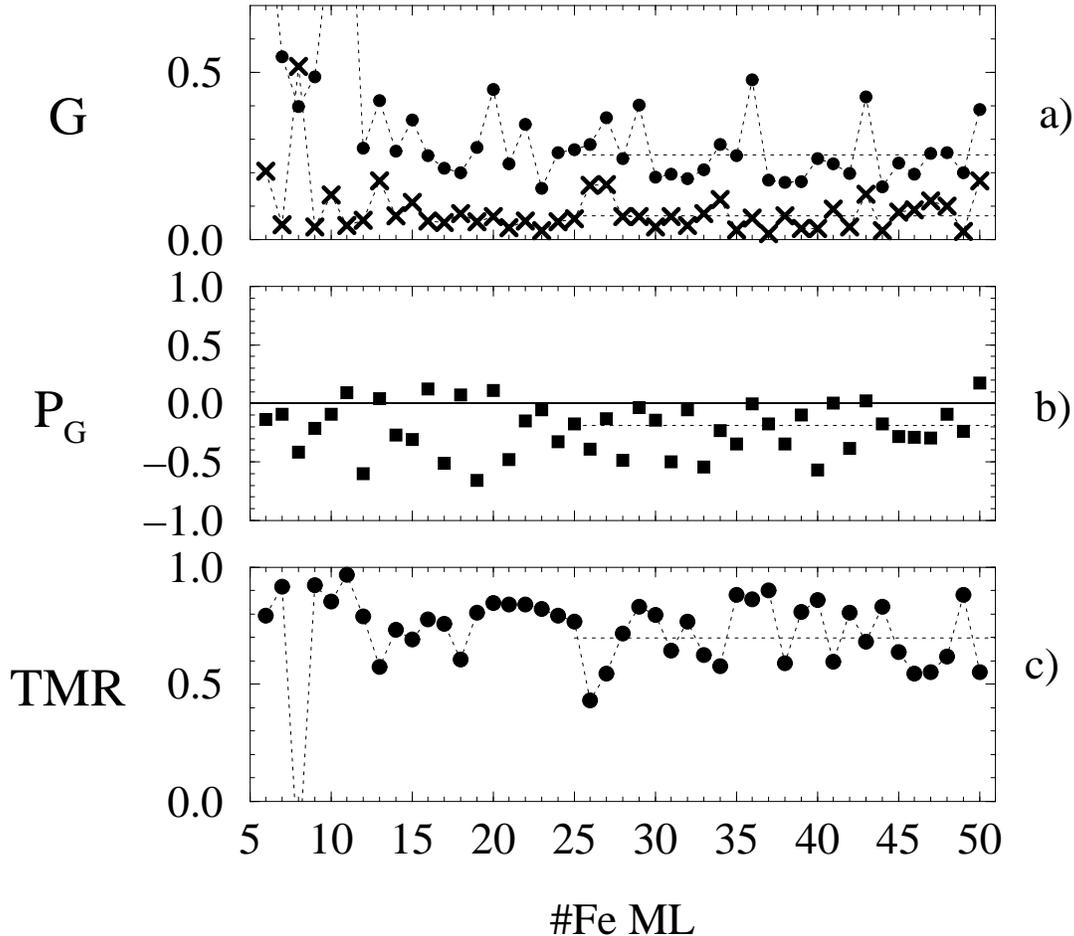}
\end{center}
 \caption{
Dependencies on the Fe layer thickness of a) conductance for P
and AP configuration, b) spin polarization $P_G$ of P conductance $G^P$, and
c) TMR ratio; the horizontal lines indicate the average
}
\end{figure}
Fig. 3a shows the corresponding conductance for P and AP configuration
in dependence on the Fe layer thickness with a barrier consiting of 4
vacuum layers (0.55 nm). In the asymptotic region (for thick Fe layers) the
conductance is a superposition of various oscillatory contributions
around a constant value of 0.256 for P and 0.077 for AP configuration,
respectively. All values of G are given in atomic units of the Rydberg
system ($G_0 = 1.752 \cdot 10^{14} \Omega^{-1} m^{-2}$).
Characteristic oscillation periods are 2.3, 3.5 and about 7.5 ML.
These periods are related to stationary points on the bulk Fe Fermi surface
and describe wave vectors at which an remarkable amount of states at $E_F$
is quantized \cite{bruno95,wildberger98}.

The Polarization P$_G$ of the conductance in P configuration
(Fig. 3b) behaves non-monotonic. The mean value in the asymptotic
region is about -0.2. This is consistent with the negative polarization
of the local density of states at the Fe surface and in the vacuum
barrier region. MacLaren et al. \cite{maclaren97} obtained a
positive polarization near 1 for a semi-infinite Fe/Vacuum/Fe junction.
The difference should first be related to the structure of the barrier consisting
of ASA potentials of constant height and secondly to the different
geometry of the leads. In our calculation the potentials are
calculated self-consistently which leads to different shapes of the
barrier potential. The barrier profile, however, can influence the
transmission coefficients drastically as it is known from free electron
models.

The TMR ratio in dependence on the Fe layer thickness is shown in Fig. 3c.
The horizontal line for thicknesses larger than 25 indicates the 
average of about 0.7. To elucidate the origin of the oscillations of 
conductance and TMR we investigated the total density of states of the
periodic system at the Fermi level as a function of the Fe layer thickness
(Fig. 4). 
\begin{figure}
\begin{center}
 \includegraphics[clip=true,width=\M]{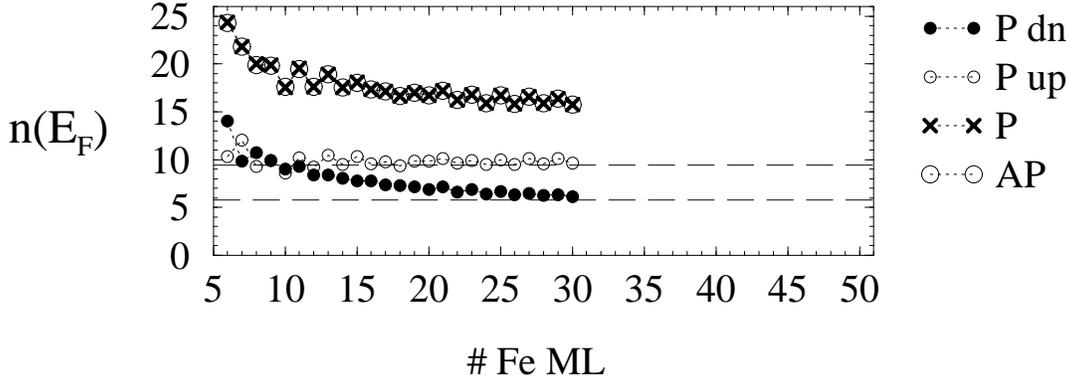}
\end{center}
 \caption{
Total DOS at the Fermi level of the Fe/Vc supercell in
dependence on Fe layer thickness for P and AP configuration at a fixed
barrier thickness of 4 ML ($0.55 nm$);
the horizontal lines indicate the Fe bulk values for both spin
directions
}
\end{figure}
Obviously, their is no difference for the density of states
in P and AP configuration of the junction, that is, TMR is not caused
by changes in the density of states. The enhancement of the
density of states due to surface resonances dominates the minority
band for thin Fe layers. For thicker Fe layers the spin-projected
density of states tends towards the Fe bulk values indicated by the 
horizontal lines. The variations of the total density for large thicknesses 
are very small and can not account for the oscillations of the conductances.
That illustrates the fact that the conductance perpendicular to the 
layers is not caused by the huge number of states at the Fermi energy
but by a few states with a large velocity in current direction.
The relative position of these states with respect to the Fermi energy 
can change drastically with the Fe layer thickness due to quantum confinement 
and consequently causes the conductance and TMR oscillations.

\section{Summary}
The ballistic conductance was calculated for an Fe/Vacuum/Fe superlattice
using the Landauer formula. 
Although we considered very large Fe
layer thicknesses up to 50 ML (about 7 nm) the conductance and the TMR ratio
still oscillates. The origin of the oscillations is the finite size
of the Fe electrodes. The calculated mean value of the conductance polarization
$P_G = -0.2$ agrees with the polarization of the local density of
states at the Fe surface.
The calculated average of the TMR ratio is 0.7.
%

%
\end{document}